\documentclass[aps,prl,twocolumn,showpacs,amsmath,amssymb,floatfix]{revtex4}
\usepackage{epsfig}
\usepackage{bm}
\def\l{\langle\!\langle}
\def\r{\rangle\!\rangle}

\begin{document}

\title{Noise-induced phase transition in the electronic Mach-Zehnder interferometer}

\author{Ivan P. Levkivskyi$^{1,2}$ and Eugene V. Sukhorukov$^1$}
\affiliation{$^1$D\'epartement de Physique Th\'eorique, Universit\'e de Gen\`eve, CH-1211 Gen\`eve 4, Switzerland}
\affiliation{$^2$Physics Department, Kyiv National University, 03022 Kyiv, Ukraine}
\date{\today}

\begin{abstract}
We consider dephasing in the electronic Mach-Zehnder  interferometer strongly coupled to 
current noise created by a voltage biased quantum point contact (QPC). 
We find the visibility of Aharonov-Bohm oscillations as a function voltage bias and express
it via the cumulant generating function of noise. In the large-bias regime,  
high-order cumulants of current add up to cancel the dilution effect of a QPC. 
This leads to an abrupt change in the dependence of the visibility on voltage bias which
occurs at the QPC's transparency $T=1/2$. Quantum fluctuations in the vicinity of this point
smear out the sharp transition.
\end{abstract}

\pacs{73.23.-b, 03.65.Yz, 85.35.Ds}

\maketitle

The effective theory of quantum Hall (QH) edge states \cite {Wen} suggests that at integer filling factors
the low-energy edge excitations are free chiral electrons. If this were the case, it would imply
that edge excitations remain coherent at long distances, and would call for various quantum information
applications. Results of tunneling spectroscopy experiments \cite{Chang} reasonably agree with the free-electron  
description of edge states. In contrast, the first experiment on Aharonov-Bohm (AB) 
oscillations of a charge current in the electronic Mach-Zehnder (MZ) interferometer \cite{firstMZ} has 
shown that the phase coherence  is strongly suppressed at energies, which are inverse proportional to the 
interferometer's size. Moreover, subsequent  experiments  \cite{Heiblum,Basel,Glattli,Litvin} have found that 
the visibility of AB oscillations as a function of voltage bias applied to the interferometer shows unusual 
lobe-type behavior, suggesting that a strong  Coulomb interaction might be responsible for dephasing of edge 
electrons.
 
Early attempts to explain the unusual AB effect in MZ interferometers 
have focused on the filling factor $\nu=1$ state, and suggested different mechanisms of dephasing, 
including the resonant interaction with a counter-propagating edge state \cite{Sukh-Che}, the dispersion 
of the Coulomb interaction potential \cite{Chalker}, and non-Gaussian noise effects \cite{Neder,Sim}. 
To date, however, all the experiments, reporting multiple side lobes in the visibility function 
of voltage bias, have been done at filling factor $\nu=2$. In one of our previous works \cite{our}, 
we have shown that in this case the long-range Coulomb interaction splits the spectrum of collective
charge excitations at the QH edge (plasmons) in two modes: a fast charge mode and a slow dipole mode. 
At low energies, only slow mode is excited at the first QPC. It carries away the electron phase information, 
but may be absorbed at the second QPC. This process partially restores the phase coherence at specific 
values of voltage bias, and generates multiple lobes in the visibility. At the same time, thanks to the chirality
of edge states, the electron transport through a single QPC is not affected by interaction. 

\begin{figure}[h]
\epsfxsize=7cm
\epsfbox{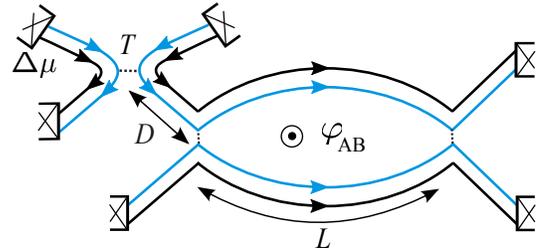}
\caption{Schematic of the electronic MZ interferometer. 
Two chiral channels are formed at the edge of a quantum Hall liquid at filling factor $\nu = 2$. 
Outer channels (shown by blue lines) are mixed at two QPCs and form an  Aharonov-Bohm loop. 
Electrons are injected into the interferometer through an additional voltage biased QPC,
which is placed at the distance $D$ from the interferometer and has transparency $T$.
} \vspace{-3mm}
\label{nu2}
\end{figure}

Importantly, the experiments \cite{Heiblum,Basel,Glattli,Litvin} can be roughly grouped in two categories 
according to whether dephasing in MZ interferometers is cause by spontaneous emission of plasmons, 
addressed earlier in 
Refs.\ \cite{Sukh-Che,Chalker,our}, or it is induced by external noise sources. In the present Letter, 
we consider the second group of experiments, where electrons are injected into a MZ interferometer via an 
additional QPC, as shown in Fig.\ \ref{nu2}. Apart from diluting the incoming electron channel, this 
additional QPC generates a partition noise \cite{Buttiker}. The MZ interferometer turns out to be strongly 
coupled to this noise, so that non-Gaussian effects, characterized by irreducible moments (cumulants) of 
the current noise, become important. We express the visibility of AB oscillations in the differential 
conductance in terms of the cumulant generating function, and find that in the limit of large voltage bias, 
all the current cumulants add up 
to cancel the dilution effect of an additional QPC. We predict, that this leads to a phase transition
at the QPC's transparency $T=1/2$, where the visibility function of voltage bias abruptly changes its behavior. 
Quantum fluctuations smear out the sharp transition in the vicinity of the critical point.

{\it Electronic Mach-Zehnder interferometer.}--- The model of a MZ interferometer, introduced earlier in 
Refs.\ \cite{Sukh-Che} and \cite{our}, is discussed here only briefly. We note, that experimentally relevant energy 
scales are very small \cite{Heiblum,Basel,Glattli,Litvin}. Therefore, it is appropriate to use an effective theory 
\cite{eff-theory} describing edge states at filling factor $\nu =2$ as collective fluctuations of the charge
density $\rho_{s\alpha}(x)$, where $\alpha= 1,2$ enumerates channels at the QH edge, and  $s = U,D$ enumerate arms of the interferometer.
The charge density fields are expressed in terms of chiral boson fields, $\phi_{s\alpha}(x)$,
satisfying the commutation relations
\begin{equation}
[\phi_{s\alpha}(x),\phi_{k\beta}(y)]=i\pi\delta_{sk}\delta_{\alpha\beta}{\rm sgn}(x-y),
\label{fields}
\end{equation}
namely, $\rho_{s\alpha}(x)=(1/2\pi)\partial_x\phi_{s\alpha}(x)$.  
The total Hamiltonian of a MZ interferometer, $\mathcal{H}=\mathcal{H}_0+\sum\nolimits_\ell(A_\ell+A_\ell^\dagger)$, 
contains a term describing edge states 
\begin{equation}
\hspace*{-1pt}\mathcal{H}_0 = \frac{1}{8\pi^2}\!\sum_{s,\alpha,\beta} \!\int\! dxdy V_{\alpha\beta}(x-y)
\partial_x\phi_{s\alpha}(x)\partial_y\phi_{s\beta}(y),
\label{hamilt}
\end{equation}
where the kernel, $V_{\alpha\beta}(x-y) = 2\pi v_F \delta_{\alpha\beta}\delta(x-y) + U_{\alpha\beta}(x-y)$,
includes a free fermion contribution with the Fermi velocity $v_F$, and the Coulomb interaction potential $U_{\alpha\beta}$.
Vertex operators $A_\ell= t_\ell \exp[i\phi_{D1}(x_\ell)-i\phi_{U1}(x_\ell)]$, $\ell = L,R$, describing electron tunneling between outer edge 
channels of the interferometer at the left and right QPC, are treated perturbatively.  
The AB phase $\varphi_{\rm AB}$ is taken into account via the relation for tunneling amplitudes,
$t_R^*t_L = |t_Rt_L|e^{i\varphi_{AB}}$.

The electron current is defined as a rate of change of the electron number in the lower arm, 
$I = i[\mathcal{H},N_D]$. To leading order in tunneling amplitudes, its average value 
is given by the linear response formula 
$\langle I\rangle = \int_{-\infty}^\infty dt\sum_{\ell\ell'} \langle[A^\dagger_\ell(t),A_{\ell'}(0)] \rangle$. 
The AB oscillations in the differential conductance $G \equiv d\langle I\rangle/d\Delta\mu$ 
are characterized by the visibility ${\cal V}_{\rm AB}(\Delta\mu) =(G_{\rm max}-G_{\rm min})/(G_{\rm max}+G_{\rm min})$.
Using the linear response formula for current, one easily finds that both the visibility and the phase shift of 
AB oscillations are expressed in terms of the same complex function \cite{our}, namely
\begin{subequations}
\begin{eqnarray}
{\cal V}_{\rm AB} = {\cal V}_0|\mathcal{I}(\Delta\mu)|, \quad \Delta\varphi_{\rm AB} = \arg \mathcal{I}(\Delta\mu),
\label{visib1}\\
\mathcal{I}(\Delta\mu) = \partial_{\Delta\mu}\int\limits_{-\infty}^\infty \frac{dt}{2\pi} K_U(L,t)K_D^*(L,t),
\label{visib2}
\end{eqnarray}
\label{visib}
\end{subequations}
where ${\cal V}_0\propto 2|t_Lt_R|/(|t_L|^2+|t_R|^2)$, and 
\begin{equation}
K_s(x,t) \propto \langle \exp[-i\phi_{s1}(x,t)]\exp[i\phi_{s1}(0,0)]\rangle
\label{corrdef} 
\end{equation}
are the electron correlation functions \cite{Giamarchi} 
at the outer channels of the interferometer. 

{\it Correlation functions and FCS.}---
The Hamiltonian (\ref{hamilt}), together with the commutation relations (\ref{fields}),
generates equations of motion for the fields $\phi_{s\alpha}$, which have to be accompanied with 
a boundary condition:
\begin{subequations}
\begin{eqnarray}
\partial_t\phi_{s\alpha}(x,t) = -\frac{1}{2\pi}\sum_\beta\int dy V_{\alpha\beta}(x-y)\partial_y\phi_{s\beta}(y,t),
\label{eoma}\\
\partial_t\phi_{s\alpha}(-D,t) = 2\pi j_{s\alpha}(t).
\label{eomb}
\end{eqnarray}
\label{eom}
\end{subequations}
We place the boundary at the point $x=-D$, where the upper outer channel is interrupted
by a QPC. 
In general, the fields $\phi_{s\alpha}$ influence fluctuations of the currents $j_{s\alpha}$ at a QPC, leading 
to the dynamical Coulomb blockade in the quantum, low-energy regime \cite{dynamicalCB}, and to cascade corrections 
to noise in the classical limit \cite{nagaev}. An important simplification in the present case arises from the fact 
that such back-action effects are absent for chiral edge states \cite{Sukh-Che,our}. As a consequence, 
in the case of $\nu=2$ the electron transport through a single QPC is not affected by interactions, which has been 
recently confirmed in the experiment \cite{Basel}. Therefore, by solving Eqs.\ (\ref{eom}), one may express 
the correlation functions of the fields $\phi_{s\alpha}$ in terms of irreducible moments (cumulants) 
of the currents, $\l j^n_{s\alpha}\r$, and equivalently, via the generator of full counting statistics (FCS) 
defined as \cite{Levitov}, 
\begin{equation}
\chi_{s\alpha}(\lambda,t)=\langle e^{i\lambda Q_{s\alpha}(t)}e^{-i\lambda Q_{s\alpha}(0)}\rangle,
\label{fcs}
\end{equation}
where $\partial^n_{i\lambda} \log(\chi_{s\alpha})/t=\l j^n_{s\alpha}\r$ in the long-time limit.  
Here, averaging is defined over free electrons, and $Q_{s\alpha}(t)=\int_{-\infty}^t dt' j_{s\alpha}(t')$.

All the interaction effects are encoded in a solution of Eq.\ (\ref{eoma}).
We assume that the Coulomb potential is screened at distances $d$, with $L\gg d \gg a$, 
where $a$ is the distance between edge channels. The screening may occur  due to the presence of
either a back gate, or a massive air bridge \cite{our}. Therefore, at low energies one can 
neglect the logarithmic dispersion of the Coulomb potential and simply write 
$U_{\alpha\beta}(x-y) = U_{\alpha\beta}\delta (x-y)$. Nevertheless, the long-range character 
of the interaction, i.e., the fact that $d\gg a$, allows one to approximate 
$U_{\alpha\beta} =\pi u$, where $u/v_F\sim\log(d/a)\gg 1$.  As a result, the spectrum of collective
charge excitations splits in two modes: a fast charged mode with the speed $u$, and a slow dipole 
mode with the speed $v \simeq v_F$.  At relevant energies, $v/L$, the charged mode is not 
excited, which leads to a universality in the electron transport predicted in Ref.\ \cite{our} and observed
in experiments \cite{Heiblum,Basel,Glattli,Litvin}. Here, taking the limit $u\to\infty$ 
simplifies the solution of Eq.\ (\ref{eoma}), and we 
obtain the result $\phi_{s1}(x,t)=-\pi[Q_{s1}(t)+Q_{s2}(t)+Q_{s1}(t_D)-Q_{s2}(t_D)]$, where $t_D = t- (x+D)/v$. 

Finally, we further assume that the noise source is located far away from the interferometer, $D\gg L$, 
which reasonably agrees with the experimental situation \cite{Heiblum,Basel,Glattli,Litvin}.
This assumption does not spoil the physics that we address, and may be relaxed later. It implies that 
all four charges in the solution for the field $\phi_{s1}(x,t)$ are uncorrelated and contribute independently
to the correlation function (\ref{corrdef}). Therefore, the correlator $K_s(x,t)$ splits in the product 
of four terms  
\begin{multline}
K_s(L,t) \propto \chi_{s1}(\pi,t)\chi_{s1}(\pi,t-L/v)\\
\times\chi_{s2}(\pi,t)\chi_{s2}(-\pi,t-L/v),
\label{cf}
\end{multline}
where we used the definition (\ref{fcs}) for the generator of FCS. 

\begin{figure}[t]
\epsfxsize=8cm
\epsfbox{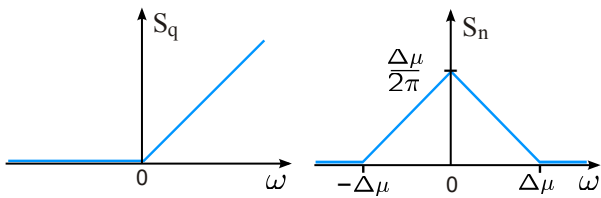}
\caption{Two spectral functions that contribute to the noise power (\ref{spectr}).
} 
\vspace{-5mm} \label{noise}
\end{figure}

{\it Gaussian noise approximation.}--- We note, that the variable $\lambda$
in the expression (\ref{cf}) plays a role of a coupling constant in the context of the noise detection physics \cite{Levitov}. 
It is typically small, so the contribution of high-order cumulants of noise to the detector signal 
is negligible \cite{Edwards}. Here, in contrast, $\lambda=\pm\pi$, implying that a MZ interferometer is strongly 
coupled to noise. Nevertheless, it is instructive, for comparison purpose, to consider Gaussian fluctuations first.
Expanding the generator (\ref{fcs}) up to second order in charge operators,
we obtain 
\begin{equation}
\log[\chi_{s\alpha}(\lambda,t)]=i\lambda \langle j_{s\alpha}\rangle t -\lambda^2J_{s\alpha}(t),
\label{logchi}
\end{equation}
where a Gaussian noise contribution is given by the integral
\begin{equation}
J_{s\alpha}(t)\equiv\frac{1}{2\pi}\int\frac{d\omega S_{s\alpha}(\omega)}{\omega^2+\eta^2}(1-e^{-i\omega t}),
\quad \eta\to 0,
\label{gaussian}
\end{equation}
and $S_{s\alpha}(\omega)=\int dt e^{i\omega t}\langle\delta j_{s\alpha}(t)\delta j_{s\alpha}(0)\rangle$
is the noise power. 

The expression (\ref{gaussian}) for the correlation function $J_{s\alpha}(t)$ is
typical in the context of the noise detection physics (see, e.g., the Ref.\ \cite{Edwards}). 
In the long-time (classical) limit, a dominant contribution to this function is linear in time:
$J_{s\alpha}(t)=(1/2)\l j^2_{s\alpha}\r|t|$, where $\l j^2_{s\alpha}\r\equiv S_{s\alpha}(0)$, 
in agreement with the definition (\ref{fcs}) of the FCS generator.
For a QPC at zero temperature, the scattering theory \cite{Buttiker} gives
\begin{equation}
S_{s\alpha}(\omega)= S_{\rm q}(\omega) + R_{s\alpha} T_{s\alpha} S_{\rm n}(\omega), 
\label{spectr}
\end{equation} 
where $S_{\rm q}(\omega) =(1/2\pi)\omega  \theta(\omega)$ is the quantum, ground-state spectral function, and 
$S_{\rm n}(\omega) =\sum_\pm S_{\rm q}(\omega\pm\Delta\mu)
-2S_{\rm q}(\omega),$ is the non-equilibrium contribution (see Fig.\ \ref{noise}). 
Note, that the noise power (\ref{spectr}) differs from the one for a non-chiral case \cite{Edwards}.

We now focus on the specific situation shown in Fig.\ \ref{nu2}, namely, we set $T_{D1}=T_{D2}=T_{U2}=1$ 
and $T_{U1}=T=1-R$. We evaluate the electron correlation function (\ref{cf}) in the upper arm of 
the MZ interferometer, using Eqs.\ (\ref{logchi}), (\ref{gaussian}) and (\ref{spectr}),
and arrive at the result
\begin{multline}
K_U(L,t)\propto \frac{\exp\{i\Delta\mu T (t-L/2v)\} }{\sqrt{t(t-L/v)}}\\
\times \exp\{-\pi^2 RT [J_{\rm n}(\Delta\mu t)+ J_{\rm n}(\Delta\mu t-\Delta\mu L/v)]\},
\label{cf-gauss}
\end{multline}
where the function $J_{\rm n}$ is given by the integral (\ref{gaussian}) with $S_{s\alpha}(\omega)$
replaced by $S_{\rm n}(\omega)$. In the expression (\ref{cf-gauss}), 
the numerator in the first term originates from the average current $T\Delta\mu/2\pi$ in (\ref{logchi}),
the denominator is the contribution of the quantum noise $S_{\rm q}(\omega)$, and 
the last term comes from the non-equilibrium noise $S_{\rm n}(\omega)$ and describes dephasing.  
The correlation function in the lower arm of the interferometer can be obtained from Eq.\ (\ref{cf-gauss})
by setting $\Delta\mu=R=0$ with the result $K_D(L,t)\propto 1/\sqrt{t(t-L/v)}$. 
Thus for a ballistic channel, and for $L=0$, the electron correlation function coincides 
with the one for free electrons. This explains the fact that in the $\nu=2$ case, the Coulomb interaction 
does not affect an electron transport through a single QPC \cite{Basel}, and justifies our approach.

Next, we use the results for correlation functions $K_s$ to evaluate the integral (\ref{visib2}).
For a large voltage bias $L\Delta\mu/v\gg 1$, we obtain 
\begin{subequations}
\begin{eqnarray}
\mathcal{I}(\Delta\mu)&\propto & E_{\rm lb}\partial_{\Delta\mu} \sin \left(\frac{\pi\Delta\mu}{E_{\rm lb}}\right)e^{-\Delta\mu/E_{\rm df}},
\label{v-integral}\\
E_{\rm lb}&=&\frac{2\pi v}{TL},\quad E_{\rm df}=\frac{4v}{\pi RTL}\,.
\label{scales}
\end{eqnarray}
\label{visib-results}
\end{subequations}
Thus the visibility ${\cal V}_{\rm AB}$, given by Eq.\ (\ref{visib1}), shows a lobe-type
behavior: It oscillates as a function of voltage bias $\Delta\mu$, vanishes at certain values of bias,
and decays. Since the function $\mathcal{I}(\Delta\mu)$ is real, the AB phase shift $\Delta\varphi_{\rm AB}$
jumps by $\pi$ at zeros of the visibility and remains constant between zeros, thus showing 
the phase rigidity \cite{Heiblum}. The distance between zeros of the visibility, $E_{\rm lb}$, 
is determined by the average current of transmitted electrons, and can be viewed 
as a ``mean-field'' contribution to the correlator (\ref{cf-gauss}). The dephasing rate, $E_{\rm df}$,
is determined by the current noise power. The ratio $2E_{\rm lb}/(\pi E_{\rm df})=R$ is given, in general,
by the Fano factor of Gaussian noise.
 
{\it Noise induced phase transition.}---
In what follows, we consider non-Gaussian noise, and show that 
the contribution of high-order cumulants of current is indeed not small. 
Note, that the ground state contribution of the current noise, $S_{\rm q}$, that 
dominates at short times, is pure Gaussian. Therefore, the denominator in the expression 
(\ref{cf-gauss}) remains unchanged. In the long time limit, the dominant
contribution to the FCS generator comes from the non-equilibrium part of noise,
$S_{\rm n}$. For a QPC, it is given by the well known expression \cite{Levitov}
for a Binomial process: $\chi_{U1}(\lambda,t)=(R+Te^{i\lambda})^N$, where $N=\Delta\mu t/2\pi$
is the number of electrons that contribute to noise.
Applying the analytical continuation $\lambda\to \pi$, we obtain 
\begin{equation}
\log[\chi_{U1}(-\pi,t)]=\frac{\Delta\mu t}{2\pi}\big[\log|T-R|+i\pi\theta(T-R)\big],
\label{logchi-qpc}
\end{equation}
where the imaginary part contributes to the effective voltage bias in the 
first term of the correlator (\ref{cf-gauss}), 
while the real part is responsible for dephasing.

\begin{figure}[h]
\epsfxsize=8.5cm
\epsfbox{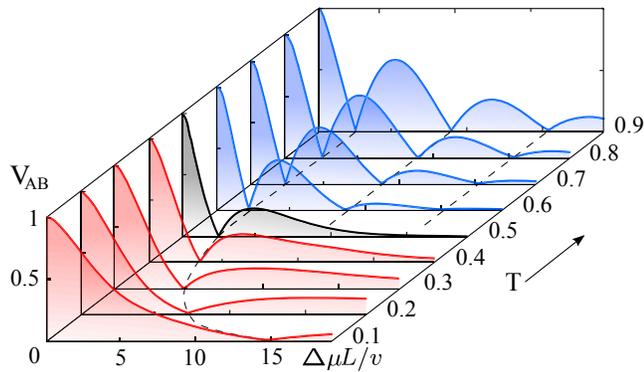}
\caption{The visibility of AB oscillations is shown as a function of the normalized
voltage bias for different transparencies of the QPC that injects electrons. 
It is evaluated numerically using the Gaussian approximation at low bias, 
and Markovian FCS at large bias. The visibility shows several lobes for $T>1/2$ (blue curves), 
while it has only one side lobe (red curves)  for $T<1/2$.  
The black curve shows the visibility at critical point of the phase transition. 
Dashed lines indicate the position of zeros.
} \vspace{-5mm}
\label{visib-shift}
\end{figure}

A remarkable property of the expression (\ref{logchi-qpc}) is that high-order cumulants 
of current add up to cancel the dilution effect of a QPC. 
Therefore, the continuous variation of the 
mean-field contribution in the correlator (\ref{cf-gauss}) is replaced with the jump in 
the voltage bias across a MZ interferometer at the point $T=1/2$. We evaluate the integral 
(\ref{visib2}) in the limit $L\Delta\mu/v\gg 1$ 
and arrive at the result (\ref{v-integral}), as in the Gaussian case, but
with new energy scales:
\begin{equation}
E_{\rm lb}=\frac{2\pi v}{L},\quad E_{\rm df}=\frac{2\pi v}{L|\log(T-R)|},\quad T>1/2.
\label{scales-fcs}
\end{equation}  
The rigidity of zeros of the visibility for $T>1/2$ is clearly 
seen in Fig.\ (\ref{visib-shift}). For $T<1/2$, the visibility may be found by taking the limit
$E_{\rm lb}\to\infty$ in the expression (\ref{v-integral}) with the result 
$\mathcal{I}(\Delta\mu)\propto  (1-\Delta\mu/E_{\rm df})e^{-\Delta\mu/E_{\rm df}}$.
Thus, the only zero of the visibility scales as $\Delta\mu=E_{\rm df}$, given 
by the expression in (\ref{scales-fcs}).
  
The behavior of the visibility of AB oscillations, shown in Fig.\ \ref{visib-shift},   
may be considered a phase transition, because strictly speaking, it arises in the classical regime,
where the number of electrons that contribute to this effect is large, $N\gg 1$. 
The transition occurs at the critical point, $\lambda=\pi$, $T=1/2$, where the moment generator 
$\chi_{U1}(\lambda,t)$ of a Binomial process vanishes, and can be viewed as a result of entanglement 
between electrons of the noise source and those that contribute to AB oscillations. 
However, quantum fluctuations of $N$ at critical point smear out the sharp transition.  
  
{\it Quantum correction at critical point.}--- 
Finding quantum corrections to the FCS of non-interacting electrons requires the evaluation of 
Fredholm determinants, which is best formulated in the wave-packet basis \cite{Levitov}. 
In the present situation a simplification arises from the fact, that in the long-time limit 
the dominant contribution to the generator (\ref{fcs}) comes from non-equilibrium electrons 
in the energy interval $\Delta\mu$. Such electrons can be viewed as a ``train'' of incoming 
wave packets $W(s_n)=\sqrt{\Delta\mu/2\pi v_F}\sin(s_n)/s_n$, where $s_n=(\Delta\mu/2) (x/v_F-t)+\pi n$, which
are normalized as $\int dx|W(s_n)|^2=1$. If electrons were transmitted through the QPC
(placed at $x=0$ for the convenience) with the probability $T$ and reflected with the probability $R=1-T$,
this would lead to a Binomial process. However, the fact that wave packets have a finite width leads to the 
small probability $P_n=\int_{-\infty}^0dxW^2(s_n)$ for electrons not to reach the QPC, which can be well
approximated with $P_n=[\pi(\Delta\mu t-2\pi n)]^{-1}$. Thus, taking into account all three possibilities,
we write the moment generating function as $\chi_{U1}(\lambda,t)=\prod_n[(1-P_n)(R+Te^{i\lambda})+P_n]$.
At critical point, $\lambda=\pi$, $T=1/2$, this gives the following result
\begin{equation}
\log[\chi_{U1}]=\sum_n\log(P_n)=-\frac{\Delta\mu t}{2\pi}[\log(\pi\Delta\mu t)-1].
\label{logchi-train}
\end{equation}
The imaginary part of $\log[\chi_{U1}]$ comes from a branch cut of the logarithm and 
grows gradually in the interval $T-R\approx 1/(2\pi^2N)$, smearing out the discontinuity
in (\ref{logchi-qpc}). Using Eq.\ (\ref{logchi-train}) we find that at critical point 
the visibility scales as ${\cal V}_{\rm AB}\propto \partial_\varepsilon 
\exp\{-\varepsilon[\log(\pi^2\varepsilon)-1]\}/\sqrt{\varepsilon}$, $\varepsilon=\Delta\mu L/2\pi v\gg 1$.
The result of a numerical evaluation is shown by black line in Fig.\ \ref{visib-shift}.

We acknowledge support from the Swiss NSF. 

\bibliographystyle{apsrev}

\begin{thebibliography}{99}

\bibitem{Wen} X.-G. Wen, {\em Quantum Field Theory of Many-Body Systems}
(Oxford University Press, Oxford, 2004).

\bibitem{Chang}
For a review, see
A.M. Chang, Rev. Mod. Phys. {\bf 75}, 1449 (2003).

\bibitem{firstMZ}
Y.\ Ji {\em et al}., Nature (London) {\bf 422}, 415 (2003).

\bibitem{Heiblum}
I. Neder {\em et al}., Phys. Rev. Lett. {\bf 96}, 016804 (2006);
I. Neder {\em et al}., Nature Physics {\bf 3}, 534 (2007).

\bibitem{Basel}
E. Bieri {\em et al.}, arXiv:cond-mat/0812.2612.

\bibitem{Glattli}
P. Roulleau {\em et al.}, Phys. Rev. B {\bf 76}, 161309(R) (2007).

\bibitem{Litvin}
L.V. Litvin {\em et al.}, Phys. Rev. B {\bf 78}, 075303 (2008).

\bibitem{Sukh-Che}
E.V. Sukhorukov, V.V. Cheianov, Phys. Rev. Lett. {\bf 99}, 156801 (2007).

\bibitem{Chalker}
J.T. Chalker, Y. Gefen, and M.Y. Veillette,
Phys. Rev. B {\bf 76}, 085320 (2007).

\bibitem{Neder} 
I. Neder, E. Ginossar, Phys. Rev. Lett. 100, 196806 (2008).

\bibitem{Sim}
S.-C. Youn, H.-W. Lee, and H.-S. Sim, Phys. Rev. Lett. {\bf 100}, 196807 (2008).

\bibitem{our}
I.P. Levkivskyi, E.V. Sukhorukov, Phys. Rev. B {\bf 78}, 045322 (2008).

\bibitem{Buttiker}
For a review, see Y.M. Blanter, M. B\"uttiker, Phys. Rep. {\bf 336}, 1 (1986).

\bibitem{eff-theory}
J. Fr\"{o}hlich, A. Zee, Nucl. Phys. B{\bf 364}, 517 (1991).

\bibitem{Giamarchi}
Th. Giamarchi, {\em Quantum Physics in One Dimension}
(Oxford University Press, Oxford, 2003).

\bibitem{dynamicalCB} For a recent experiment, see
C. Altimiras {\em et al.},
Phys. Rev. Lett. {\bf 99}, 256805 (2007).

\bibitem{nagaev}
K. E. Nagaev, Phys. Rev. B {\bf 66}, 075334 (2002).

\bibitem{Levitov} L.S. Levitov, H. Lee, and G.B. Lesovik, J.\ Math.\ 
Phys. {\bf 37}, 4845 (1996). 

\bibitem{Edwards}
E. V. Sukhorukov, and J. Edwards, Phys. Rev. B {\bf 78}, 035332 (2008). 



\end{thebibliography}

\end{document}